\documentclass[useAMS,usenatbib]{mn2e}
\usepackage[T1]{fontenc}
\usepackage{graphics,graphicx,epsfig}
\usepackage{amssymb,amsmath,amsfonts,amssymb,subfigure}
\usepackage{mathrsfs}
\title[Flat Central Density Profile and Constant DM Surface Density in Galaxies from Scalar Field Dark Matter]{Flat 
Central Density Profile and Constant DM Surface Density in Galaxies from Scalar Field Dark Matter}
\author[Victor H. Robles and T. Matos]{Victor H. Robles\thanks{E-mail:
vrobles@fis.cinvestav.mx} and T. Matos\thanks{E-mail:
tmatos@fis.cinvestav.mx}\\
Departamento de F\'isica,Centro de Investigaci\'on y de Estudios Avanzados del IPN, AP 14-740, 0700  D.F., M\'exico\\}
\begin{document}

\maketitle

\label{firstpage}

\begin{abstract}
The scalar field dark matter (SFDM) model proposes that galaxies form by condensation of a scalar field (SF) very early in the 
universe forming Bose-Einstein Condensates (BEC) drops, i.e., in this model haloes of galaxies are gigantic drops of SF. 
Here big structures form like in the LCDM model, by hierarchy, thus all the predictions of the LCDM model at big scales are 
reproduced by SFDM. This model predicts that all galaxies must be very similar and exist for bigger redshifts than in the LCDM model. 
In this work we show that BEC dark matter haloes fit high-resolution rotation curves of a sample of thirteen low surface
brightness galaxies. We compare our fits to those obtained using a Navarro-Frenk-White and Pseudo-Isothermal (PI) profiles and 
found a better agreement with the SFDM and PI profiles. The mean value of the logarithmic inner density slopes is 
$\alpha = - $0.27 $\pm$ 0.18. As a second result we find a natural way to define the core radius with the advantage of being 
model-independent. Using this new definition in the BEC density profile we find that the recent observation of the constant dark 
matter central surface density can be reproduced. We conclude that in light of the difficulties that the standard model is currently
facing the SFDM model can be a worthy alternative to keep exploring further.

\end{abstract}


\section{Introduction}
In the $\Lambda$ Cold Dark Matter ($\Lambda$CDM or LCDM) model, known as the standard model of cosmology, 
the formation of structures in the universe is through a hierarchical process of growth of structures, meaning that small structures,
like small haloes of galaxies merge to form bigger ones, like galaxy clusters and superclusters haloes. 
In this picture, the universe contains around 96 per cent of an unknown 
form of energy that is usually called dark matter (DM) and dark energy.
The $\Lambda$CDM model can successfully describe cosmological observations 
such as the large scale distribution of galaxies, the temperature variations in the 
cosmic microwave background radiation and the recent acceleration of the universe
\citep{a43,a44,a45,a46}.

However, recent observations in far and nearby galaxies have shown that the model 
faces serious conflicts when trying to explain the galaxy formation at small scales \citep{a1,a32}.
For instance, in the $\Lambda$CDM simulations the halos present rising densities towards the central region 
behaving as $\rho \sim r^{-1}$ well within 1 kpc \citep{a2}. On the other hand various 
observations suggest that the rotation curves are more consistent with a constant central density
\citep{a3,a4}, this is most commonly known as the cusp/core problem \citep{a4}.

Studying a wide range of galaxies of different morphologies and with magnitudes in the interval $-22 \leq M_{B} \leq -8$ 
\citet{a5} fit their rotation curves (RC) using a Burkert profile for the DM \citep{a6} 
and found that 
\begin{equation}
\log (\mu_{0}/M_{\odot}pc^{-2})= 2.15 \pm 0.2 \label{eq:logmu}
\end{equation}
remains approximately constant, where 
\begin{equation}
\mu_{0} = \rho_{0}r_{0} \label{eq:mu}
\end{equation}
with $\rho_{0}$ the central DM
density and $r_{0}$ the core radius. Similar results where found in \citet{a7,a8}. 
Exploring further the constant value of $\mu_{0}$ for the DM, \citet{a9} found that within $r_{0}$ 
the DM central surface density in terms of the mass inside it ($M_{< r_{0}}$) is $<\Sigma>_{0,DM} = M_{< r_{0}} / \pi r^{2}_{0}
\approx 72^{+42}_{-27} M_{\odot} pc^{-2}$,the gravitational acceleration due to DM felt by a test particle at the radius $r_{0}$
was found to be 
\begin{equation}
g_{DM}(r_{0})=G \pi <\Sigma>_{0,DM} = 3.2^{+1.8}_{-1.2}\times 10^{-9} cm s^{-2}, \label{eq:gDM}
\end{equation}
additionally they reported the acceleration due to the luminous matter at
$r_{0}$ to be $g_{bar}(r_{0})= 5.7^{+3.8}_{-2.8}\times 10^{-10} cm s^{-2}$.

In the $\Lambda$CDM model the galaxies have evolved through numerous mergers and grown in different environments, the 
star formation and basic properties of the galaxies are not expected to be a common factor among them. Therefore 
explaining both the constancy of $\mu_{0}$ and the core in the central regions of galaxies seems very unlikely in this model.

These problems can be used to test alternative DM models. There are other models that do not include DM but instead 
modify the Newtonian force law \textbf{F}= m\textbf{a}, one of them is called MOND and was proposed by
Milgrom \citep{a10,a11}, in this modification Newton's law reads
\textbf{F} = m\textbf{a}$\mu (a/a_{0})$, where the fixed acceleration scale $a_{0}$ divides 
the Newtonian and MONDian regimes, for $x<<1$ $\mu(x)= x$ we have the MONDian regime and for $x>>1$ we recover the usual Newtonian 
acceleration, the value of $a_{0} \sim 1.2 \times 10^{-10} m/s^{2}$.

Lately, the scalar field dark matter (SFDM) model has received much attention.
When the scalar field contains a self-interaction this model is also called the Bose-Einstein Condensate (BEC) dark matter 
model, both names are used in the literature and are interchangeable.The main idea is simple \citep{a36}.
The SFDM model proposes that galaxies form by condensation of a scalar field (SF) with an ultra-light mass of the order of 
$m_{\phi} \sim 10^{-22}$eV. Therefore, when we mention the 
SFDM or BEC model in this paper we are describing a scalar field that condensates somehow and becomes Bose Einstein 
Condensate dark matter. From this mass it follows that the critical temperature of condensation 
$T_c\sim1/m_{\phi}^{5/3}\sim$TeV is very high, thus, they form Bose-Einstein Condensates (BEC) drops very early in the universe. 
It has been proposed that these drops are the haloes of galaxies (see \citet{a37}), $i.e.$, that haloes are gigantic
drops of SF. On the other side, big structures form like in the LCDM model, by hierarchy \citep{a37,a38}, thus, all 
predictions of the LCDM model at big scales are reproduced by SFDM. In other words, in the SFDM model the haloes of galaxies 
do not form hierarchically, they from at the same time and in the same way when the universe reaches the critical temperature
of condensation of the SF, in a similar way as water drops form in the clouds. From this it follows that all galaxies must be
very similar because they formed in the same manner and at the same moment. Therefore, from this model we have to expect that 
there exist well formed galaxy haloes at bigger redshifts than in the LCDM model.  In this model the scalar particles with 
this small mass are such that their wave properties avoid the cusp and reduce the high number of small satellites \citep{a14}
which is another problem that is still present in the $\Lambda$CDM model \citep{a1,a15}.Summarizing, it is remarkable that
with only one free parameter, the ultra-light scalar field mass ($m_{\phi} \sim 10^{-22}$eV), the SFDM model fits:
\begin{enumerate}
\item The evolution of the cosmological densities \citep{a39}.

\item  The rotation curves of big galaxies \citep{a19,a40} and LSB galaxies. 

\item With this mass, the critical mass of collapse for a real scalar field is just $10^{12}\,M_{\odot}$, i.e., the one 
observed in galaxy haloes \citep{a41}.
 
\item The scalar field has a natural cut off, thus the substructure in clusters of galaxies is avoided naturally. 
With a scalar field mass of $m_\phi\sim10^{-22}$eV the amount of substructure is compatible with the observed one \citep{a37,a38}.

\item We expect that SFDM forms galaxies earlier than the cold dark matter model, because they form BECs at a 
critical temperature $T_c >> $ TeV. So if SFDM is right, we have to see big galaxies at big redshifts with similar features.  

\item And recently it has been demonstrated that SFDM haloes maintain satellite galaxies going around big galaxies 
for enough time to explain the existence of old stars in the satellites, provided that the mass of the SF is just 
$m_{\phi}\sim10^{-22}$eV  \citep{a42}.
\end{enumerate}
The idea was first considered by \citet{a12,a13} and independently introduced by \citet{a36}.
In the BEC model, DM halos can be described in the non-relativistic regime, where DM halos can 
be seen as a Newtonian gas. If we consider a SF self-interaction, we need to add a quartic term to the SF potential,
in this case the equation of state of the SF is that of a polytope of index n=1 \citep{a38,a19}. 
Different issues of BEC DM halos and the cosmological behavior of the BEC model have been studied in \citet{a16,a17,a18,a19,a20,a35}. 

It is a fact that any model trying to become a serious alternative to $\Lambda$CDM has to succeed not only in reproducing 
observations in which the standard model fails but also has to keep the solid description at large scale.
For this reason, our aim in this work is to test the BEC model with the two observations mentioned above, 
the cusp/core problem \citep{a18} and the constant DM central surface density. In order to do this, 
we used the Thomas-Fermi approximation and a static BEC DM halo to fit rotation curves of a set of galaxies. 
However, so far there was no comparison between the density profile 
and the data, in this work we fill this blank by fitting rotation curves of 13 high resolution 
low surface brightness (LSB) galaxies and additionally compare the fits to two
characteristic density profiles 1) the cuspy Navarro-Frenk-White (NFW) profile that results from N-body simulations 
using $\Lambda$CDM and 2) the Pseudo Isothermal (PI) core profile. The comparison allow us to show 
that the model is in general agreement with the data and with a core in the central region. 
For our second result we found that the meaning of a core is somewhat ambiguous. In order to
clarify the meaning and unify the description, we propose a new 
definition for the core and core radius that allow us to decide when a density profile is cusp or core.
Using this definition in the BEC model discussed above we find that the BEC model can reproduced the constant 
value of $\mu_{0}$ and as a crosscheck we used the PI profile and find our results to be in very good agreement with observations.
This argues in favor of the model and our definition.

In section 2, we describe the density profiles we use to fit the rotation curves and 
we give our definition of the core and core radius. In section 3, we fit the galaxy data and obtain 
the fitting parameters and the core radius for the PI and BEC density profiles. In section 4, we 
discuss our results and in section 5 we give our conclusions.

\section[]{Dark matter halo density profiles}
In this section we provide the dark matter profiles that will be used in the analysis. 
In the last part of the section we briefly describe the usual meaning of the core and 
establish a new definition for it.

\subsection{BEC profile}
The case in which the dark matter is in the form of a static Bose-Einstein condensate 
and the number of DM particles in the ground state is very large was considered in \citet{a18}.
Following this paper and assuming the Thomas-Fermi approximation \citep{a21} which 
neglects the anisotropic pressure terms that are relevant only in the boundary of the condensate,
the system of equations describing the static BEC in a gravitational potential V is given by
\begin{equation}
\nabla p \biggl( \dfrac{ \rho}{m} \biggr) = - \rho \nabla V \label{eq:BEC1}
\end{equation}
\begin{equation}
 \nabla^{2} V = 4 \pi G \rho, \label{eq:BEC2}
\end{equation}
with the following equation of state 
\begin{equation}
p(\rho) = U_{0} \rho^{2},
\end{equation}
where $ U_{0} = \dfrac{2 \pi \hbar^{2} a}{m^{3}}$, $\rho$ is the mass density of the static BEC 
configuration and $p$ is the pressure, as we are considering zero temperature $p$ is not a thermal pressure
but instead it is produced by the strong repulsive interaction between the ground state bosons. 
Assuming spherical symmetry and denoting $R$ as the radius at which the pressure and density are zero, 
the density profile takes the form \citep{a18} 
\begin{equation}
 \rho_{B} (r) = \rho^{B}_{0} \dfrac{\sin( k r)}{k r} \label{eq:BEC}
\end{equation}
where $k = \sqrt{G m^{3} / \hbar^{2} a}=\pi/R$ and $ \rho^{B}_{0}=\rho_{B}(0)$ 
is the BEC central density, $m$ is the mass of the DM particle and $a$ is the 
scattering lenght. The mass at the radius $r$ is given by 
\begin{equation}
m(r)= \frac{4 \pi \rho^{B}_{0}}{k^{2}} r \biggl( \dfrac{\sin (kr)}{kr } - \cos(kr) \biggr ), \label{eq:BECm}
\end{equation}
from here the tangential velocity $V_{B}$ of a test particle at a distance $r$, is  
\begin{equation}
V^{2}_{B}(r) = \dfrac{4 \pi G \rho^{B}_{0} }{k^{2}} \biggl( \dfrac{\sin (kr)}{kr } - \cos(kr) \biggr ). \label{eq:BECV}
\end{equation}
The logaritmic slope of a density profile is defined as 
\begin{equation}
\alpha = \dfrac{d (\log \ \rho)}{d (\log \ r)} 
\end{equation}
using (7) in (10) it is obtained \citep{a47} 
\begin{equation}
\alpha(r)= - \biggl [1- \dfrac{\pi r}{R} \cot \biggl(\dfrac{\pi r}{R} \biggr)  \biggr].
\end{equation}
Additionaly, the logarithmic slope of the rotation curve is defined \citep{a47} by
\begin{equation}
\beta = \dfrac{d (\log \ V)}{d (\log \ r)} 
\end{equation}
from (9) we get 
\begin{equation}
\beta = -\dfrac{1}{2} \biggl [1- \dfrac{( \pi r / R)^{2}}{1- (\pi r /R ) \cot ( \pi r /R )}  \biggr].
\end{equation}

\subsection{Pseudo Isothermal profile}

All the empirical core profiles that exist in the literature fit two parameters, a scale
radius and a scale density. A characteristic profile of this type is
\begin{equation}
\rho_{PI} = \dfrac{\rho^{PI}_{0}}{1 \ + \ (r/R_{c})^{2}},\label{eq:PI}
\end{equation} 
this is the PI profile \citep{a22}. Here $R_{c}$ is the scale radius and $\rho^{PI}_{0}$ is the central 
density. The rotation curve is 
\begin{equation}
V(r)_{PI} = \sqrt{4 \pi G \rho^{PI}_{0} R^{2}_{c} \biggl ( 1-\frac{R_{c}}{r} \arctan \biggl ( \dfrac{r}{R_{c}} \biggr) \biggr ) }.
\end{equation}

\subsection{Navarro-Frenk-White profile}

The NFW profile emerges from numerical simulations that use only CDM and are based on the $\Lambda$CDM model \citep{a23,a24,a25}. 
In addition to this, we have chosen this profile because it is representative of what is called 
the cuspy behavior ($\alpha \approx -1$) in the center of galaxies due to DM. 
The NFW density profile \citep{a25} and the rotation curve are given respectively by
\begin{equation}
\rho_{NFW}(r) = \dfrac{\rho_{i}}{(r/Rs)(1 \ + \ r/R_{s})^{2}} 
 \end{equation}
\begin{equation}
V_{NFW}(r)= \sqrt{4 \pi G \rho_{i} R^{3}_{s}} \sqrt{ \frac{1}{r} \biggl [ ln \biggl (1 \ + \ \dfrac{r}{R_{s}} \biggr ) - 
\dfrac{ r/R_{s}}{1\ + \ r/R_{s} } \biggr ]},
\end{equation}
$\rho_{i}$ is related to the density of the universe at the moment the halo collapsed and $R_{s}^{2}$ 
is a characteristic radius.
\subsection{Meaning of the core radius and cusp/core discrepancy}

In the large scale simulations that use collissionless cold dark matter the inner 
region of DM halos show a density distribution described by a power law $ \rho \sim r^{\alpha}$ with $\alpha \approx -1$,
such behavour is what is now called a cusp. On the other hand, observations mainly in dwarf and LSB galaxies 
seem to prefer a central density going as $ \rho \sim r^{0}$. 
This discrepancy between observation and the CDM model receives the name of cusp/core problem. 
Among the empirical profiles most frequently used to describe the constant density behavior 
in these galaxies are the PI \citep{a22}, the isothermal \citep{a26} and the Burkert
profile \citep{a6}. Even though their behavior is similar in the central region and is specified by the 
central density fitting parameter, their second parameter called the core radius  
does not represent the same idea. For instance, in the PI profile (eq.(\ref{eq:PI})) we see that the core radius will be the 
distance in which the density is half the central density. For the Burkert profile the core radius $R^{burk}_{c}$ 
will be when $\rho^{burk}(R^{burk}_{c}) = \rho^{burk}_{0}/4$ and for an isothermal profile (I) \citep{a8} 
$\rho^{I}(R^{I}_{c}) = \rho^{I}_{0}/2^{3/2}$. Hence, we see an ambiguity in the meaning of the core radius, 
they get the same name but the interpretation depends on the profile. If we want 
to compare the central density of LSB galaxies with that of NFW, it usually suffice to have a 
qualitative comparison, so far this is what we have been doing by fitting 
empirical profiles. However, high resolution rotation curves demand a more quantitative 
comparison. Indeed, if we want to test models by fitting RCs we have to know the specific 
meaning and size of the core, then we will be able to tell if a model is consistent with 
a cusp or not by making a direct comparison with the data. 

For this reason we ask ourselves the question: \textsl{What and where is the core?} 
To solve the previous ambiguity and to unify the concept for future comparisons, we found that a good 
definition for the $core$ is a region where the density profile presents logarithmic slopes $\alpha \geq -1$ and 
the core radius will be the radius at which the core begins, that is to say, for radius smaller 
than the core radius we will have $\alpha \geq -1$, this means that its value $r'$ is determined by the equation
\begin{equation}
\alpha(r') = -1
\end{equation}
The advantages of this definition are that the interpretation is independent of the profile 
chosen (also notice that it applies to the total density profile and is 
not restricted to that of DM) and in virtue of the same definition 
we can directly tell if a DM model profile is cored or cuspy.  
With our new definition the specific distance at which the core radius occurs still 
depends on the profile chosen but now the physical interpretation is only one. 
In the following when we refer to both the core and core radius we adopt the previous interpretation. 

Applying the definition to (\ref{eq:BEC}) we get the core radius for the BEC profile $R_{B}$ and 
for comparison we use (\ref{eq:PI}) because in turns out that the parameter $R_{c}$ 
corresponds to the core radius as defined above. Finally, fitting the NFW profile 
provides a direct comparison between a cusp and core and hence to the cusp/core problem.
\begin{figure}
\begin{tabular}{cc}
\includegraphics{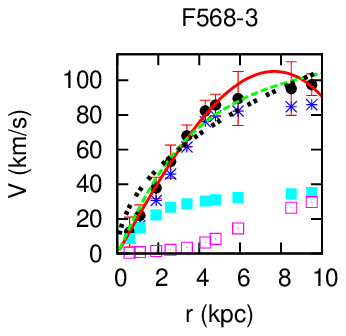} &
\includegraphics{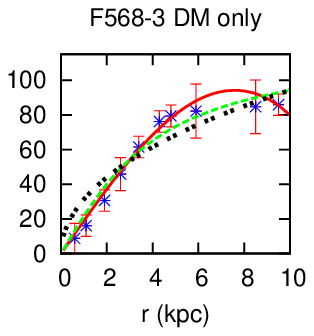} \\
\includegraphics{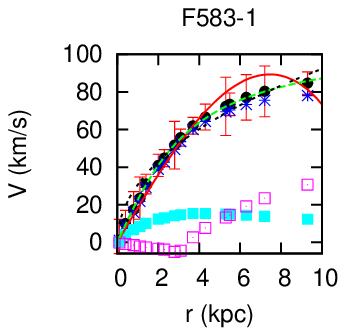} & 
\includegraphics{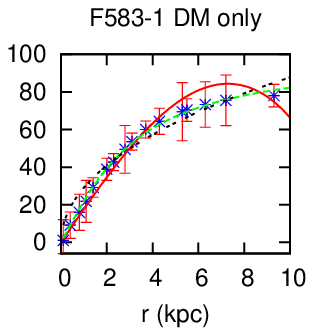} \\
\end{tabular}
  \caption{Contribution of the baryons to the rotation curve for F568-3 and F583-1.
We denote observed data by black dots with error bars, dark matter with blue asterisks,
the disk with cyan squares and the the gas with magenta squared boxes. We fit the figures on the left  
asumming the minimun disk hypothesis while for the ones on the right we substract in quadrature the baryons and fit only the dark matter.
We notice that the barionic component is not dominant in the outer regions and that the difference in the fits is barely noticeable.}
\end{figure}

\begin{figure}
\centering
 \includegraphics{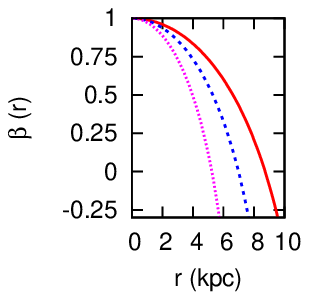}
  \includegraphics{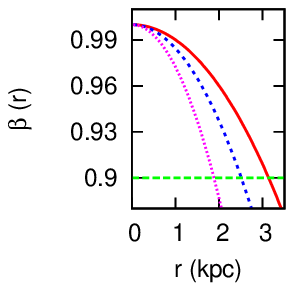}
  \caption{Logarithmic slope of the rotation curve ($\beta$) for three values of $R$. From left to right 
 $R$ = 6,8,10. The plots on the left show a common behavior always reaching zero before $R$. The 
figure on the right shows the region where a linear behavior in the RC speed is still valid, the (green) horizontal 
line bounds this region and the radius in which this boundary is reached was found to be $r\approx 0.31R$ in each curve.}
\end{figure}

\begin{figure*}
\begin{minipage}{170mm}
\begin{tabular}{@{}lll@{}}
 \includegraphics{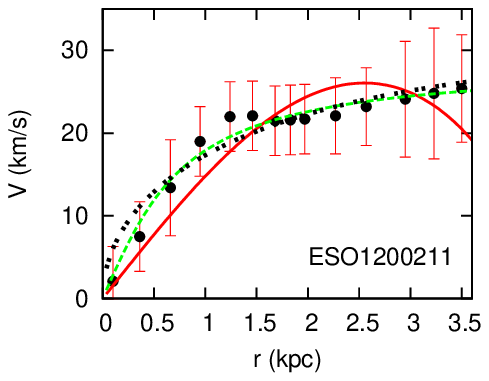} &
 \includegraphics{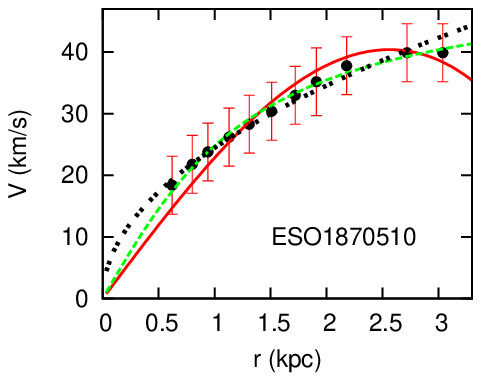} &
 \includegraphics{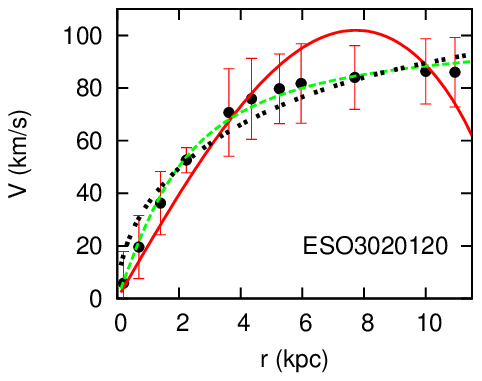} \\
\includegraphics{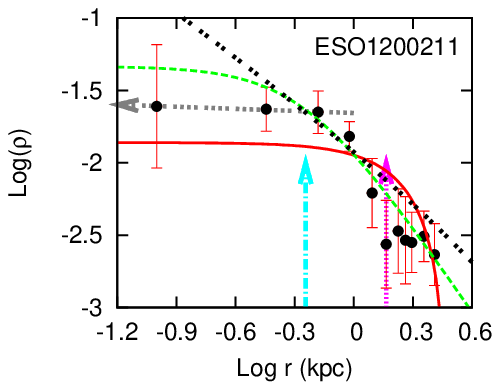} &
\includegraphics{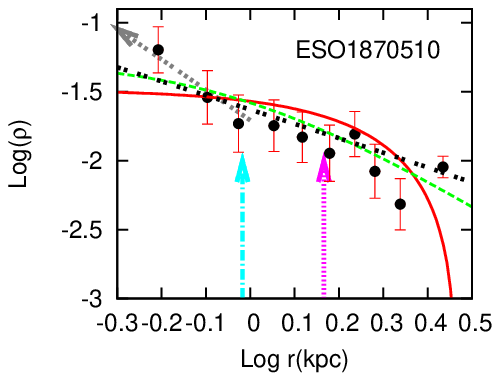} &
\includegraphics{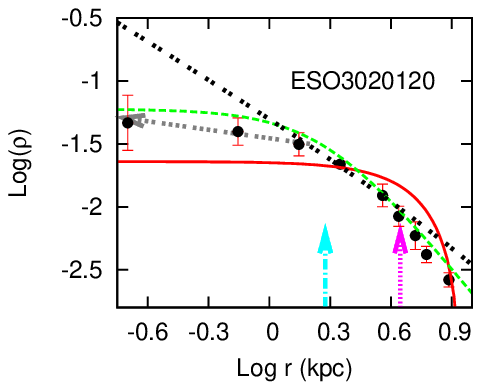} \\
\includegraphics{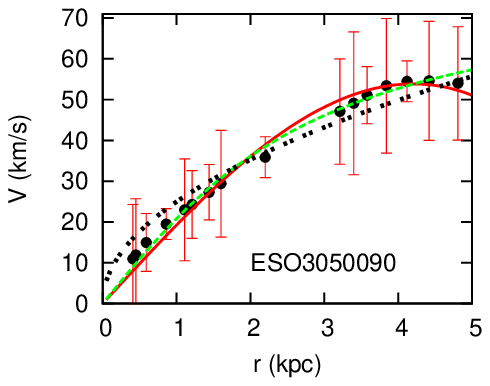} &
\includegraphics{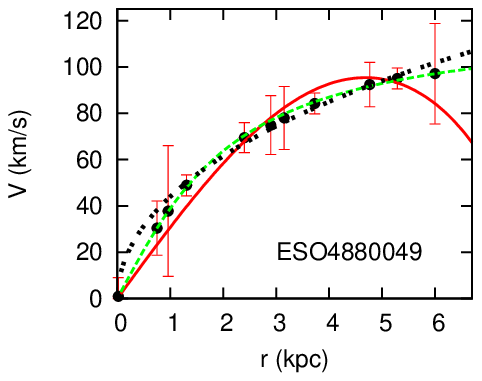} &
\includegraphics{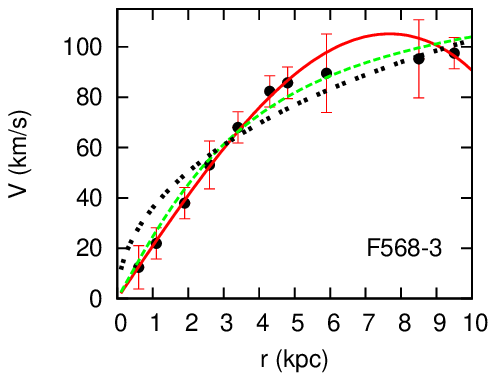} \\
\includegraphics{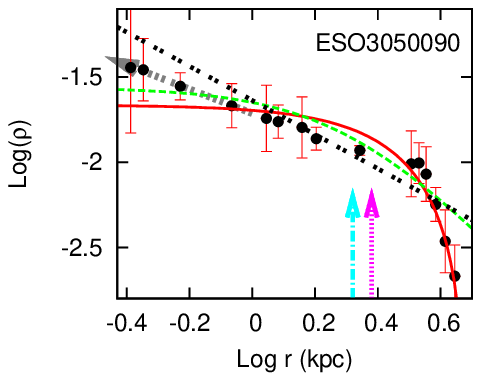} &
\includegraphics{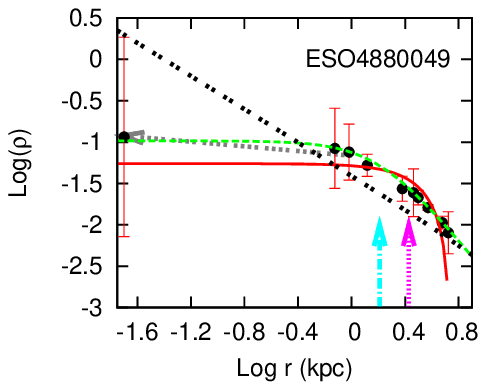} &
\includegraphics{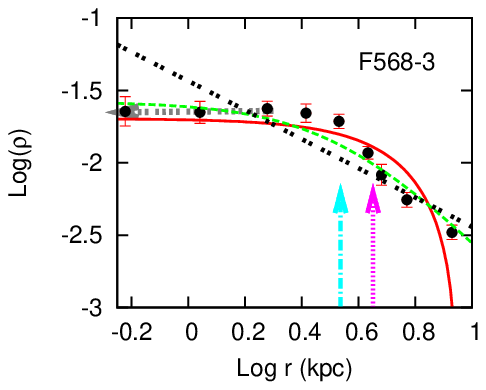} \\
\includegraphics{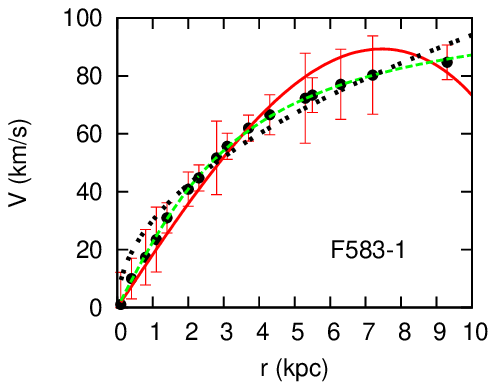} & 
\includegraphics{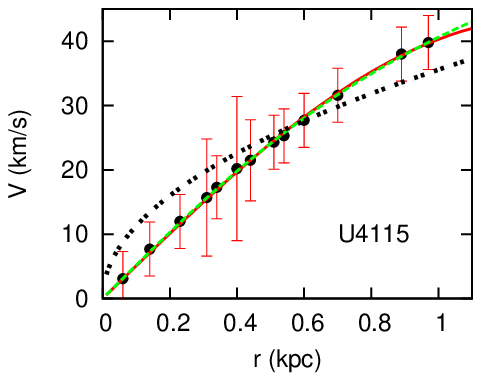} &
\includegraphics{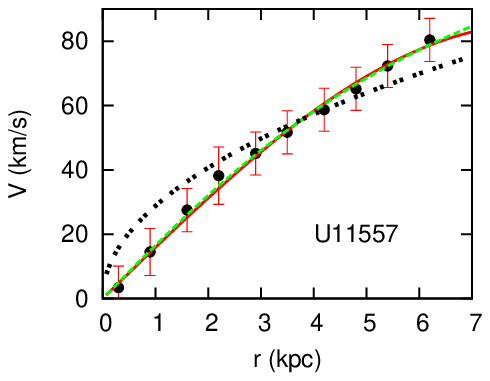} \\
\includegraphics{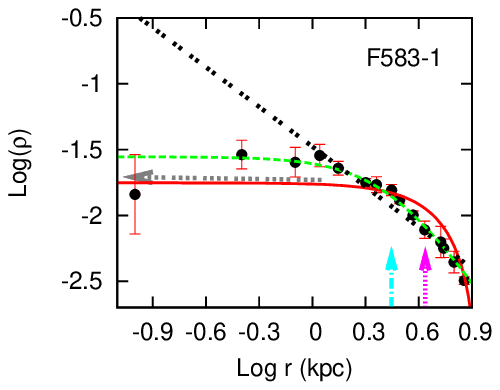} &
\includegraphics{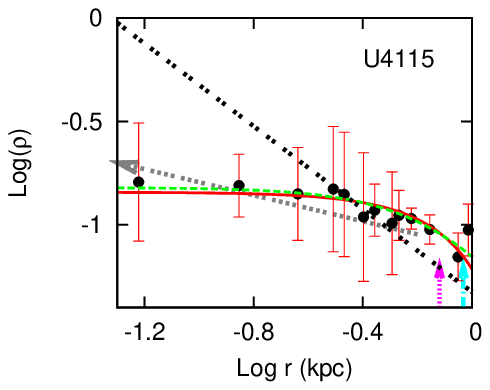} &
\includegraphics{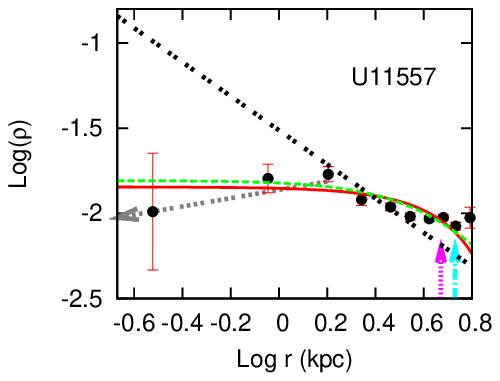} \\
\end{tabular}
\end{minipage}
\end{figure*}

\begin{figure*}
\begin{minipage}{170mm}
\begin{tabular}{@{}lll@{}}
\includegraphics{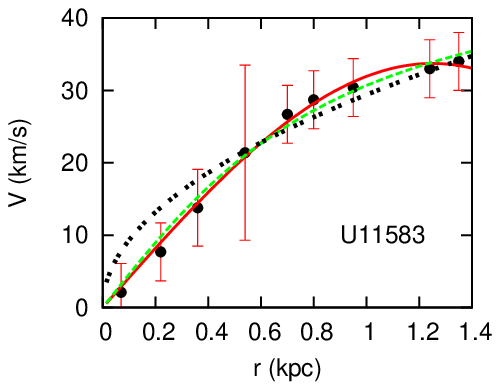} &
\includegraphics{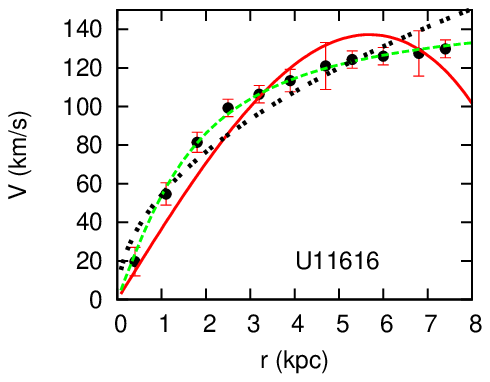} &
\includegraphics{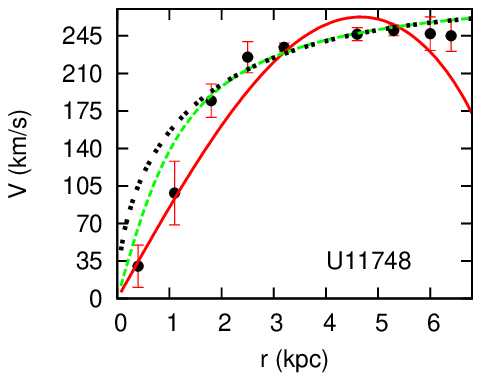} \\
\includegraphics{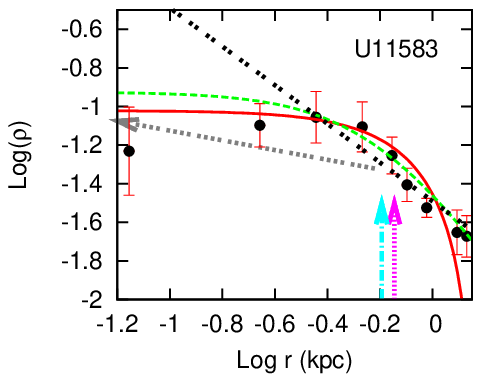} &
\includegraphics{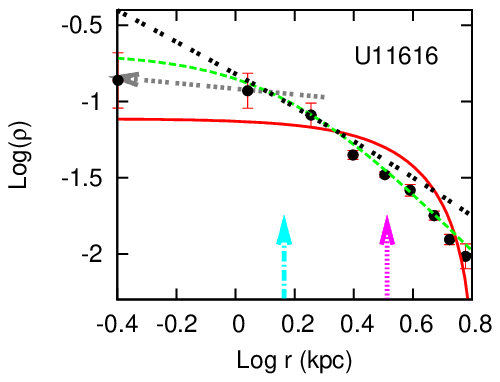} &
\includegraphics{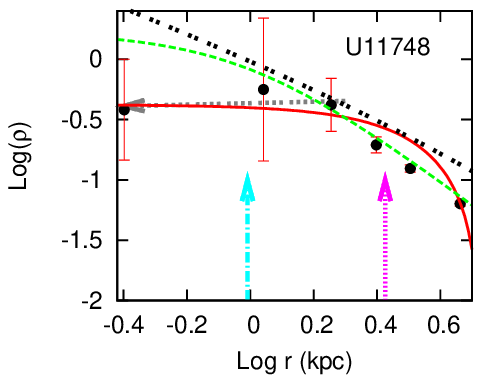} \\
\includegraphics{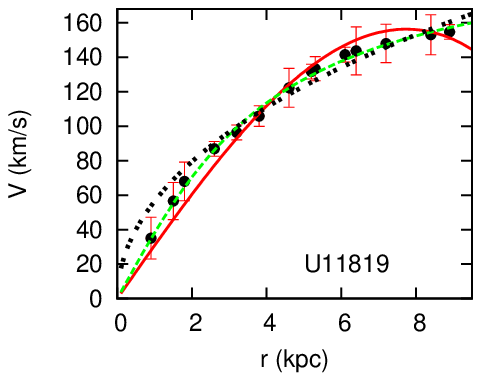} &
\includegraphics{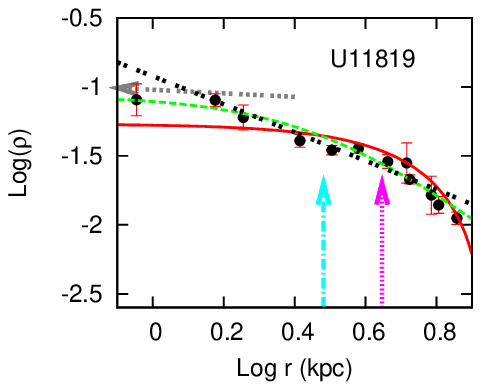} & \\
\end{tabular}
  \caption{Observed LSB galaxy rotation curves and density profiles with the 
best halo fits. Below each RC is its density profile along with the fits. 
Shown are the PI (green dashed-line); the BEC (red solid line) and 
NFW (black dobble-dotted line) DM halo profiles, the observational data is drawn with error bars. 
The gray arrow denotes the best fit to the data within $R_{1}$ and the vertical arrows 
denote the PI (blue) and BEC (magenta) core radius.}
\end{minipage}
\end{figure*}

\section[]{Fits and data}

We see from (\ref{eq:BEC}) that the BEC model satisfies $\rho \sim r^{0}$ near the origin, but a priori 
this does not imply consistency with observed RCs. 
Therefore, we fit the profiles in section 2 to thirteen high resolution observed RCs of a sample of LSB galaxies.
The RCs were taken from a subsample of \citet{a27}, we chose galaxies that 
have at least 3 values within $\sim 1$ kpc, not presenting bulbs and the quality in the RC in H$\alpha$ 
is good as defined in \citet{a28}. The RCs in this work omit 
galaxies presenting high asymmetries and included in the error bars are experimental
errors in the velocity measurement, inclination and small asymmetries.
Because the DM is the dominant mass component for these galaxies 
we adopt the minimum disk hypothesis which neglects baryon contribution to the observed RC. 
In order to show that in LSB and dwarf galaxies neglecting the effect of baryons is a good hypothesis, we include in 
Fig.1 two representative examples (F568-3 and F583-1). For these galaxies we plot the contribution 
of the gas, disk and the dark matter separately. We did the fitting first considering the total 
contribution and then using only DM (marked as 14 and 15 in Table 1 and 2). We found no substantial difference
in our values, which can be seen from our results in Table 1 and 2. As the other galaxies belong to the category of
DM dominated galaxies as other authors have shown \citep{a3,a27}, neglecting baryons in our analyses will not 
modify substantially our results.
   
As the difference between a core and a cusp is most notable only for data values inside 1 kpc
and given that in the interval $\sim$ 1 to 10 kpc the slopes of core and cusp profiles are very similar, 
which can lead to the wrong conclusion that cuspy halos are consistent with observations, we 
determined the logarithmic slope and the uncertainty following \citet{a27} with 
the difference that we fit only the data within 1 kpc and that there is no need of an uncertain ``break radius''. 

In Table 1 we list the fitting parameters of the profiles of section 2, we also include the values
of the logarithmic slope and its uncertainty, the value $R_{1}$ denotes the nearest radius to 1kpc 
where a data point is given. We obtain $\alpha$ by fitting values inside $R_{1}$, and we also 
report the core radius for the BEC profile $R_{B}$ in order to compare it with $R_{c}$.
In Table 2 we report both the value of eq. (\ref{eq:gDM}) for the BEC profile and the logarithm of eq. (\ref{eq:mu}) 
for PI and BEC profiles. 

In Fig. 3, we show our fits to the RC data and the density profiles, also shown are 
the core radius in the BEC (magenta) and in PI (blue) profiles. The gray arrow is the fit
that determines $\alpha$, the size of the arrow denotes the fitted region and is bounded by $R_{1}$.

\begin{figure}
\centering
 \includegraphics{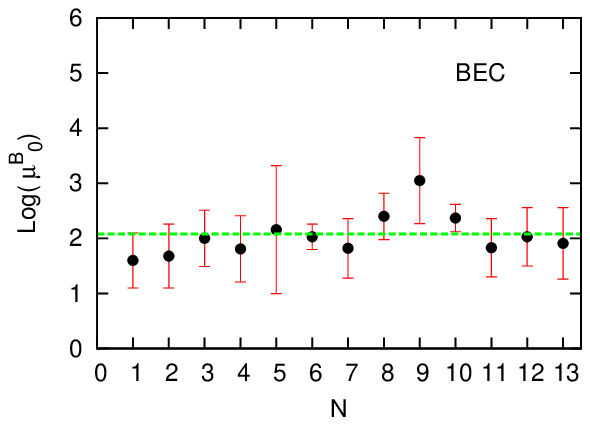}
 \includegraphics{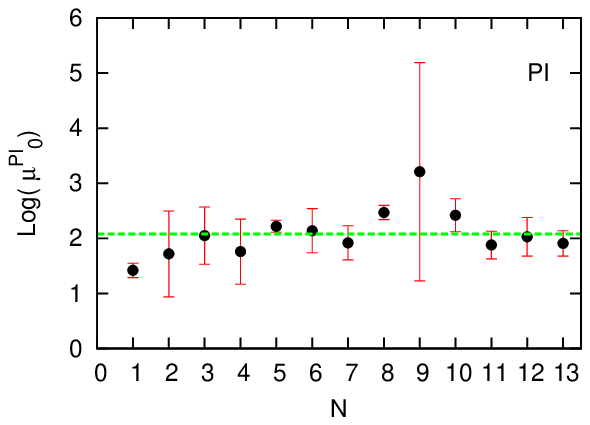}
  \caption{Plot of log $\big( \mu^{B}_{0}/M_{\odot}pc^{-2} \big)$ and 
 log $\big( \mu^{PI}_{0}/M_{\odot}pc^{-2} \big)$ for each galaxy. N denotes the 
galaxy according to Table 1. Here we observe that these values remain 
approximately constant in both profiles, this also serves as a 
crosscheck for our definition of $R_{B}$ in the BEC profile.
The green dashed-line represents the mean values given in (\ref{eq:muB}) and (\ref{eq:muPI}).}
\end{figure}

\begin{table*}
\centering
 \begin{minipage}{145mm}
  \caption{Best-fit parameters of the profiles in section 2.}
   \subtable[]{
  \begin{tabular}{@{}ccrrccccc@{}}
  \hline
  \hline 
   Label & Galaxy & $\rho^{PI}_{0}$ & $R_{c}$ & $\chi_{PI}^{2}$ & $\rho^{B}_{0}$ & R & 
   $R_{B}$ & $\chi_{B}^{2}$ \\
    & & ($M_{\odot}pc^{-3})$ & (kpc) & & $(M_{\odot}pc^{-3})$ & (kpc) & 
   (kpc) \\ 
 \hline
1 & ESO1200211 & 0.0464 & 0.57 & 1.45 & 0.0138 & 2.92 & 1.46 & 0.44 \\
2 & ESO1870510 & 0.0548 & 0.96 & 0.63 & 0.0329 & 2.93 & 1.465 & 0.2 \\
3 & ESO3020120 & 0.0598 & 1.89 & 0.03 & 0.0229 & 8.95 & 4.425 & 0.92 \\
4 & ESO3050090 & 0.0276 & 2.09 & 0.04 & 0.0217 & 4.81 & 3.04 & 0.1 \\
5 & ESO4880049 & 0.1035 & 1.62 & 0.99 & 0.0549 & 5.36 & 2.68 & 0.79 \\
6 & U4115 & 0.1514  & 0.93 & 0.10 & 0.1438 & 1.52 & 0.76 & 0.15 \\
7 & U11557 & 0.0156 & 5.37 & 0.08 & 0.0143 & 9.38 & 4.69 & 0.07 \\
8 & U11611 & 0.2065  & 1.46 & 0.14 & 0.0771 & 6.51 & 3.225 & 5.24 \\
9 & U11748 & 1.678  & 0.98 & 1.43 & 0.4205 & 5.33 & 2.665 & 4.37 \\
10 & U11819 & 0.0869 & 3.03 & 0.20 & 0.0539 & 8.86 & 4.43 & 1.09 \\
11 & U11583 & 0.119 & 0.64 & 0.12 & 0.0953 & 1.43 & 0.715 & 0.84 \\
12 & F568-3 & 0.0361 & 3.01 & 0.53 & 0.0248 & 8.78 & 4.39 & 0.11 \\
13 & F583-1 & 0.0317 & 2.6 & 0.54 & 0.019 & 8.53 & 4.26 & 0.48 \\
14 & F568-3 (DM) & 0.0264 & 3.44 & 0.84 & 0.0202 & 8.96 & 4.48 & 0.25 \\
15 & F583-1 (DM) & 0.0279 & 2.79 & 0.58 & 0.0177 & 8.66 & 4.33 & 0.37 \\
\hline
\end{tabular}
 \label{tab:firsttable}
}
\subtable[]{        
  \begin{tabular}{@{}lcrrcccc@{}}
  \hline
  \hline 
   Label & Galaxy & $\rho_{i}$ & $R_{s}$ & $\chi_{NFW}^{2}$ & $\alpha$ 
   & $\Delta \alpha$ & $R_{1}$ \\
   & & ($\times 10^{-3}M_{\odot}pc^{-3}$) & $(kpc)$ & & 
   & & (kpc) \\
 \hline
1 & ESO1200211 & 2.45 & 5.7 & 0.24 & -0.04 & 0.53 & 0.95  \\
2 & ESO1870510 & 0.761 & 31.82 & 0.05 & -1.09 & 0.76 & 1.13 \\
3 & ESO3020120 & 2.65 & 19.72 & 0.32 & -0.2 & 0.16 & 1.4  \\
4 & ESO3050090 & 0.0328 & 705.67 & 0.22 & -0.7 & 0.03 & 1.1 \\
5 & ESO4880049 & 1.42 & 52.27 & 0.16 & -0.09 & 0.06 & 0.96 \\
6 & U4115 & 0.139  & 341.74 & 0.78 & -0.27 & 0.24 & 0.89 \\
7 & U11557 & 0.0108 & 2849.65 & 1.43 & 0.2 & 0.31 & 0.9 \\
8 & U11611 & 11.59  & 14 & 1.77 & -0.15 & 0.31 & 1.1 \\
9 & U11748 & 204.58  & 5.53 & 3.41 & -0.38 & 0.11 & 1.1 \\
10 & U11819 & 1.19 & 101.24 & 1.07 & -0.64 & 0.13 & 2.5 \\
11 & U11583 & 0.136 & 238.568 & 0.81 & -0.2 & 0.24 & 0.95 \\
12 & F568-3 & 0.378 & 120.78 & 2.36 & 0.03 & 0.02 & 1.9 \\
13 & F583-1 & 0.345 & 102.349 & 0.55 & -0.03 & 0.04 & 1.1 \\
14 & F568-3 (DM) & 0.0715 & 515.68 & 2.87 & 0.34 & 0.27 & 1.9 \\
15 & F583-1 (DM) & 0.329 & 102.349 & 0.58 & 0.02 & 0.07 & 1.1 \\
\hline
\end{tabular}
       \label{tab:secondtable}
}
\end{minipage}
\end{table*}

\begin{table*}
\centering
 \begin{minipage}{110mm}
  \caption{Derived quantities from the parameters in Table 1.}
  \begin{tabular}{@{}ccrrc@{}}
  \hline
  \hline 
   Label & Galaxy & log $\mu^{PI}_{0}$ %
  \footnote{Both $\mu^{PI}_{0}$ and $\mu^{B}_{0}$ units are $M_{\odot}/pc^{2}$}
   & $log \mu^{B}_{0}$ & $g^{B}_{DM}$ \\
   & & & & $(\times 10^{-9}cms^{-2})$  \\ 
 \hline
1 & ESO1200211 & 1.42 $\pm$ 0.13 & 1.60 $\pm$ 0.50 & 0.908 \\
2 & ESO1870510 & 1.72 $\pm$ 0.78 & 1.68 $\pm$ 0.58 & 2.15 \\
3 & ESO3020120 & 2.05 $\pm$ 0.52 & 2.00 $\pm$ 0.51 & 2.17 \\
4 & ESO3050090 & 1.76 $\pm$ 0.59 & 1.81 $\pm$ 0.60 & 4.58 \\
5 & ESO4880049 & 2.22 $\pm$ 0.11 & 2.16 $\pm$ 1.16 & 6.63 \\
6 & U4115 & 2.14 $\pm$ 0.40 & 2.03 $\pm$ 0.23 & 4.93 \\
7 & U11557 & 1.92 $\pm$ 0.31 & 1.82 $\pm$ 0.54 & 3.03 \\
8 & U11611 & 2.97 $\pm$ 0.13 & 2.40 $\pm$ 0.42 & 11.3 \\
9 & U11748 & 3.21 $\pm$ 1.98 & 3.05 $\pm$ 0.78 & 50.4 \\
10 & U11819 & 2.42 $\pm$ 0.30 & 2.37 $\pm$ 0.25 & 10.8 \\
11 & U11583 & 1.88 $\pm$ 0.25 & 1.83 $\pm$ 0.53 & 3.08 \\
12 & F568-3 & 2.03 $\pm$ 0.35 & 2.03 $\pm$ 0.53 & 4.91 \\
13 & F583-1 & 1.91 $\pm$ 0.23 & 1.91 $\pm$ 0.65 & 3.66 \\
14 & F568-3 (DM) & 1.95 $\pm$ 0.51 & 1.95 $\pm$ 0.68 & 4.09 \\
15 & F583-1 (DM) & 1.89 $\pm$ 0.13 & 1.88 $\pm$ 0.59 & 3.45 \\
\hline
\end{tabular}
\end{minipage}
\end{table*}

\section[]{Discussion}

In Fig. 2 we plot $\beta$ using three different values of $R$. From this figure we see a common behavior. 
We have that $\beta$ is a decreasing function of r and is zero before $R$, which tell us that (9) always reaches a maximun before $R$.
We also notice that there is a region in which $V_{B} \sim r$, we can take this region to be when $0.9\leq \beta \leq 1 $.
If we use $\beta=0.9$ as an upper bound for the region in which the linear behavior ($V_{B} \sim r$) remains valid,
we obtain an upper bound radius of  $r\approx 0.31 R$. This means that for values of $r \leq 0.31R$  we expect $V_{B} \sim r$.
The latter can be used as a test to the BEC model by fitting the RCs and veryfing this solid-body behavior within the mentioned
region.
The fits of the RCs in Fig. 3 prove that the solid-body like behavior characterized by a linear increase of the velocity 
in the central region is satisfied by the BEC model, in fact, it is more consistent with the core PI and BEC profiles 
than the cuspy NFW. 

If we now turn to the density profiles, our fits within $R_{1}$ give an average value of $\alpha$=$-$0.27 $\pm$ 0.18 consistent 
with those obtained in \citet{a27} $\alpha$=$-$0.2 $\pm$ 0.2 and with 
$\alpha$=$-$0.29 $\pm$ 0.07 reported by \citet{a29} analyzing 7 THINGS dwarf galaxies. 
The case of ESO1870510 might be considered to be consistent with NFW profile, however 
it is the innermost value that considerably decreases $\alpha$, being an irregular galaxy 
more central data near the innermost region is required to discard the possibility of any violent event that 
might have caused such a slope value.

The density profiles corresponding to the RCs fits are also shown in Fig.3 for each galaxy. We see that  
the BEC fits slightly deviates for the farthest data points as a result of the finite size of the
radius $R$ that is fixed by the same data. This discrepancy is due to the fact that the halo might be more extended
than the value $R$. As a matter of a fact, the more extended the ``flat'' outer region in the RCs
the more conspicuous the discrepancy. 
The main reason of this comes from Fig.2 where we infer that the rotation 
curve speed always presents a maximum value followed by a continuos decrease, which means that to avoid the 
mentioned discrepancy we need that the BEC rotation curve profile remains approximately constant after its maximum.
From Fig.1 we see that the total rotation curve is dominated by the dark matter contribution, specially in the outer regions.
Hence, unless the baryons become the dominant component in the outer regions, which does not seem to be observed, 
it is unlikely that adding the barionic contribution to the RCs in our galaxies will solve the discrepancy.

Some solutions to keep the BEC rotation curve constant after its maximum include finite temperature 
corrections to (\ref{eq:PI}) \citep{a30}, this suffice to alleviate the latter problem in LSB galaxies and dwarfs but 
not for bigger galaxies. Other authors proposed including vortex lattices \citep{a31,a33} and adding more nodes \citep{a12,a13}
in the solution of system (\ref{eq:BEC1}) and (\ref{eq:BEC2}). Nevertheless, it can be shown \citep{Guzman:2003kt} 
that a systems of many nodes is unstable, therefore so far no final conclusion has been reached.

When comparing the BEC and PI core radius we find a general difference of $\sim$ 2 kpc, the core size 
in the PI profile is approximately 2 kpc smaller than the BEC core size, but the PI central density is larger.
In U4115, U11557 and U11583 both profiles are very similar which results in a similar core and 
central density values, this can also be taken as a consistency check for our core definition. 

Comparing the values of $R_{B}$ in Table 1 we did not find a tendency to a common value. 
Assuming that the core radius determines the transition where the DM distribution changes from the outer 
region to the inner constant central density, the lack of a unique value means that there is not 
a common radius at which this transition takes place.  

For our second test we use $R_{B}$ to calculate (\ref{eq:mu}). We have already seen that $R_{c}$ and  
$R_{B}$ are generally different and $R_{B}$ is not a fit parameter. Hence \textsl{a priori} $R_{B}$
is not expected to correlate with $\rho^{B}_{0}$. However, with the values of Table 1 we obtain 
\begin{equation}
\log (\mu^{B}_{0}/ M_{\odot} pc^{-2}) = \log \rho^{B}_{0} R_{B} = 2.05 \pm 0.56 \label{eq:muB}
\end{equation}
\begin{equation}
\log (\mu^{PI}_{0}/ M_{\odot} pc^{-2}) = \log \rho^{PI}_{0} R_{c} = 2.08 \pm 0.46 \label{eq:muPI}
\end{equation}
for the average values in the BEC and PI profiles respectively. We see the excellent agreement of 
(\ref{eq:muB}) with (\ref{eq:muPI}) that 
was used as a crosscheck and with (\ref{eq:logmu}) in which a much bigger sample was used.  
The agreement has shown that the BEC model is capable of reproducing the constancy 
of the value $\mu_{0}$, something that because of the cuspy nature is not possible in the NFW profile.

In Fig. 4 we plot the above values for each galaxy. We define the DM central surface density (mentioned
in the introduction) for the BEC profile by
\begin{equation}
<\Sigma>^{B}_{0,DM} = M_{< R_{B}} / \pi r^{2}_{B}, 
\end{equation}
where $M_{< R_{B}}$ is obtained from (\ref{eq:BECm}) evaluated at $R_{B}$.
From Table 2 we see that for U11748 the value $ log \mu^{B,PI}_{0}$ is considerably above the rest and 
with the largest uncertainty. For this reason, in the following analysis we omit both, this value and 
the smallest one that corresponds to ESO1200211. Doing this we get an average value of eq. (19) given by
$<<\Sigma>^{B}_{0,DM}> \approx$ 191.35 $M_{\odot}pc^{-2}$, 
and for the acceleration felt by a test particle located in $R_{B}$ due to DM only we 
have $g_{DM}(R_{B}) \approx $ 5.2 $\times 10^{-9} cm s^{-2}$ broadly consistent with (\ref{eq:gDM}).

The fact that all galaxies present approximately the same order of magnitude in $g_{DM}(R_{B})$ 
might suggest that $R_{B}$ represents more than a transition towards a constant density, it can give us 
information about the close relation between DM and the baryons. Moreover, in view of the lack 
of a unique core radius, we can interpret the transition in DM distribution as an effect 
of crossing a certain acceleration scale instead of a radial length scale. Such interpretation 
reminds us that given in MOND but with the big difference that the acceleration scale found is 
for DM and is not a postulate of the model. 

To determine which interpretation causes the transition whether an acceleration scale or a length scale
we will need to study the properties of larger samples of galaxies observed with the new telescopes.

\section{Conclusions}
In this paper we find that the BEC model gives a constant density profile that is consistent
with RCs of dark matter dominated galaxies. The profile is as good as one of the most
frequently used empirical core profiles but with the advantage of coming from 
a solid theoretical frame. We fit data within 1 kpc and found a logarithmic slope 
$\alpha$=$-$0.27 $\pm$ 0.18 in perfect agreement with a core.
It is important to notice that the cusp in the central regions is not a prediction that 
comes from first principles in the CDM model, it is a property that is derived by fitting 
simulations that use only DM. 

We established the ambiguity present in the usual interpretation of the core radius, 
we proposed a new definition for the core and core radius that takes away the ambiguity 
and that has a clear meaning that allows for a definite distinction when a density profile is core or cusp. 

Using our definition we find the core radius in the BEC profile to be in most cases over 2 kpc
bigger than the core radius in the PI profile. We have assumed the DM particles are bosons and 
that a great number of them is in the ground state in the form of a condensate. This led us to good results for 
our sample of galaxies, but it might be necessary to consider more than these simple hypotheses.

As a second result and direct consequence of our core definition, we were able to obtain the constant value of $\mu_{0}$
which is proportional to the central surface density. This result is one of several conflicts 
that jeopardize the current standard cosmological model.

If we continue to observe even more galaxies with a core behavior, this model can be a 
good alternative to $\Lambda$CDM.

\section*{Acknowledgments}
This work was partially supported by CONACyT M\'exico under grants CB-2009-01, no. 132400 and I0101/131/07 C-234/07 of 
the Instituto Avanzado de Cosmolog\'ia (IAC) collaboration (http://www.iac.edu.mx/).

\end{document}